# Lightweight Lexical Test Prioritization for Immediate Feedback

Toni Mattis[a] and Robert Hirschfeld[a]

a Hasso Plattner Institute, University of Potsdam, Germany

**Abstract** The practice of unit testing enables programmers to obtain automated feedback on whether a currently edited program is consistent with the expectations specified in test cases. Feedback is most valuable when it happens immediately, as defects can be corrected instantly before they become harder to fix. With growing and longer running test suites, however, feedback is obtained less frequently and lags behind program changes.

The objective of test prioritization is to rank tests so that defects, if present, are found either as early as possible or with the least costs. While there are numerous static approaches that output a ranking of tests solely based on the current version of a program, we focus on change-based test prioritization, which recommends tests that likely fail in response to the most recent program change. The canonical approach relies on coverage data and prioritizes tests that cover the changed region, but obtaining and updating coverage data is costly. More recently, information retrieval techniques that exploit overlapping vocabulary between change and tests have proven to be powerful, yet lightweight.

In this work, we demonstrate the capabilities of information retrieval for prioritizing tests in dynamic programming languages using Python as example. We discuss and measure previously understudied variation points, including how contextual information around a program change can be used, and design alternatives to the widespread *TF-IDF* retrieval model tailored to retrieving failing tests.

To obtain program changes with associated test failures, we designed a tool that generates a large set of faulty changes from version history along with their test results. Using this data set, we compared existing and new lexical prioritization strategies using four open-source Python projects, showing large improvements over untreated and random test orders and results consistent with related work in statically typed languages.

We conclude that lightweight IR-based prioritization strategies are effective tools to predict failing tests in the absence of coverage data or when static analysis is intractable like in dynamic languages. This knowledge can benefit both individual programmers that rely on fast feedback, as well as operators of continuous integration infrastructure, where resources can be freed sooner by detecting defects earlier in the build cycle.

## 1 Introduction

Immediacy and continuity of feedback are desirable during programming activities. Automated tests, often manifesting themselves as regression tests or *unit tests*, are a best practice to receive feedback on whether the test authors' expectations are consistent with the implementation at hand. With growing test suites and, consequentially, longer execution times, feedback becomes less immediate and is obtained less frequently because of that. As a result, the benefits of frequent testing wane, among them the cost benefits of early-caught errors, and the psychological benefits of perceiving causality between change and test failure.

**Test Prioritization** Widespread unit test runners, such as *JUnit* or *PyTest* run tests in the order in which they appear throughout the module or directory tree of the program's sources, relying on manually (de-)selecting individual modules or subdirectories if programmers need to focus on relevant tests.

The goal of *test prioritization* is to choose a better order in which tests are executed—in our case, the test that is most likely to fail should run first. To this end, several heuristics for estimating the fault-detection capability of a test have been proposed. Common criteria are statement/branch coverage, coverage of valuable or risky features, or requirement coverage. The task of a prioritization strategy is to schedule tests in a way that maximizes these criteria as early as possible.

While ranking tests does not speed up the test suite as a whole, it reduces the time programmers wait for test *failures* (if any are present) and increases the chance of fault detection under time constraints. Examples where this can be helpful include:

1. Programming environments that run tests on every change but are not given sufficient time to run the full test suite between subsequent edits.
2. Live programming environments [13] that create the impression of editing a running program and rely on immediate feedback as part of their experience.
3. Continuous integration servers, where a "red" build needs to be discovered and resolved quickly, and stopping a red build earlier frees resources for the next build or saves energy.
4. Continuous delivery infrastructures with multiple stages, where the most frequently updated stage is tested against a small, but fast subset of tests, while more stable deployments are exercised with larger and larger test suites the closer they move to production.

**Change-based Test Prioritization** Our focus on accelerating feedback while programming demands that we set our scope to approaches targeting the *most recent change* to a program. Our objective will be to rank tests according to their likelihood of detecting a recently introduced defect, and do so fast enough that prioritization itself does not significantly increase testing overhead.

Prioritizing tests based on the vocabulary they share with the most recent change is an approximate, but fast strategy to discover relevant tests. State of the art in lexical test prioritization [18] follows a standard information retrieval (IR) approach, where lexical features in each test are Term Frequency–Inverse Document Frequency (TF-IDF)-weighted, scored against a query computed from the last change, and returned in descending order with respect to this similarity score.

**Problem** As of now, studying classical IR based on TF-IDF has been restricted to statically typed languages (primarily Java), where the verbosity of type information provides plenty of comparable lexical features and alternative strategies based on static analysis provide competitive performance. Neither of these is available in a dynamically typed language like Python. Moreover, TF-IDF was intended to be used on collections of natural language documents, missing a wide range of properties inherent to program code.

**Approach** In this work, we outline how information retrieval techniques are used to prioritize unit tests, and discuss possible variations when dealing with programs rather than natural language documents. We particularly focus on the TF-IDF weighting scheme and propose alternatives that directly learn the predictive power of individual terms from previous test failures.

As a way to generate sufficient data for a study, we develop a tool to generate faulty changes with associated test runs, and evaluate both classical and newly proposed prioritization techniques on four popular and well-tested Python projects. We observe that all lexical prioritization models under consideration improve the ranking of tests significantly over a randomized or untreated order, up to a degree that can be considered *immediate* feedback. Alternatives to TF-IDF perform slightly better than the original lexical model in plain ranking performance.

In our outlook we focus on possible directions of improvement of the techniques themselves and future applications of such a model.

## 2 Lexical Test Prioritization

We are reviewing the approach of ranking tests based to the vocabulary shared with a recent change to the program. This heuristic builds on the assumptions that tests often refer to the tested entities by their name (e.g. calling a method, instantiating a class) and programmers choose similar names to refer to a concept used by both tested and testing code. An example is given in table 1. Finding data based on shared vocabulary is an IR problem and we discuss background, state-of-the-art, and tradeoffs involved in IR-based test prioritization in this section.

In contrast to Java-based literature, which often refers to a test case as a subclass of *JUnit*'s `TestCase`, we refer to the concepts *test*, *test case*, and *test method* interchangeably, designating a single method/function that performs one or more *assertions*, has a name beginning with `test` and resides in a file within a `test(s)` directory.

**Listing 1** Example of a change in production code and a test. Both share the words `user` and `admin`. According to this heuristic, the test has high fault-detection potential given the change. Generic words like the keyword `def` carry little meaning and can be excluded or discounted.

**(a)** Change (in unix diff notation) of Python web server code

```
@@-42,1 +42,2 @@
- return resource

+ if session.user.is_admin():
+   return resource
```

**(b)** Unit test checking whether a web resource is secure

```
def test_admin_access():
    assert get(...).status == 403
    with login(self.admin_user):
        assert get(...).status == 200
```

## 2.1 State of the Art

**Features** State of the art in IR-based test prioritization relies on extracting variables, methods, classes, comments, and other programmatic concepts affected by a change, normalizing these *features* (e.g. by splitting camel case identifiers, stemming, etc.), and assigning TF-IDF weights to them.

Given a change to the code base, a query vector can be computed from the features occurring in edited lines of code. The query vector is compared to each test's feature vector using a scoring function.

In this section, we shortly describe TF-IDF, scoring functions, and the widely used Okapi Best Match 25 (BM25) retrieval model.

**TF-IDF** An important part of IR is a function which estimates how relevant a word is within its respective context (often a document). The idea behind the TF-IDF scheme is to emphasize words that are locally frequent but globally rare, because local occurrence can mean that a document is about that specific concept, while global occurrence hints at common words not tied to any specific topic.

The TF-IDF mechanism constructs per-document (in our case, per-*test*) *vectors* of the size of our vocabulary, i.e., one vector dimension per unique word. The weight in each word dimension is proportional to how often it appears within that document, called Term Frequency (TF), multiplied with a factor that discounts words that occur in a large proportion of documents by multiplying with the feature's Inverse Document Frequency (IDF). The weight of a non-occurring feature can be implicitly regarded as 0, hence the vectors tend to be sparse.

Among the numerous variations of this scheme are adjustments by document length such that a feature occurring more often due to being in a longer block of test code is not at an unfair advantage.

**Ranking** A query is answered by computing a *score* of how relevant each document is to the query by summing over the TF-IDF weights of all terms in that document which happen to appear in the query. Typically the documents with the top-$n$ scores would be retrieved. In test prioritization, query vectors are computed from the lexical features that were modified or close to the modified code, and the ranking of (all) tests determines their prioritized order of execution in response to that change.

**BM25** The Okapi BM25 model is a widespread variation [15] of TF-IDF ranking to compute a document $d$'s score $S(q,d)$ with respect to a query $q$:

$$S(q,d) = \sum_{f \in q} \text{IDF}(f) \cdot \frac{\text{TF}(f,d)(k_1+1)}{\text{TF}(f,d) + k_1\left(1-b+b\left(|d|/\hat{d}\right)\right)} \quad (1)$$

Here, $\text{IDF}(f)$ of a feature $f$ is defined as $\text{IDF}(f) = log\frac{N-n_f+0.5}{n_f+0.5}$ with $N$ as the total number of documents and $n_f$ the number of documents containing feature $f$. $\text{TF}(f,d)$ is the number of times feature $f$ appears in document $d$, $|d|$ is the document length and $\hat{d}$ the average document length. $k_1$ is a free parameter that determines how fast term frequencies saturate, and the value of $b$ determines how much the frequency is scaled with respect to the document length.

### 2.2 Trade-offs and Variations

Using IR for test prioritization involves a number of trade-offs and design decisions grounded in the fact that programs have a different structure than natural language documents. So far, few of them have been explicitly addressed.

**Relevance** Current information retrieval models often rely on a variation of TF-IDF. This weighting scheme judges the relevance of a word with respect to a document based on local versus global distribution. This encodes assumptions based on the statistical properties of natural language and might not necessarily reflect those of source code.

The relevance of an identifier in source code can depend on its syntactic role, e.g. a class name is likely more relevant than a temporary variable. Related work has shown that differentiating between features from comments, class or method names, temporary variables, etc. can improve retrieval performance in fault localization tasks [17]. However, as shown by the *REPiR* project [18], test prioritization does not improve significantly.

Instead of judging the relevance of words based on their distribution or syntactical role, we propose a third interpretation of relevance: weighting terms according to their empirically observed predictive power. This lowers the influence of words that occur independently of failing tests and heightens the influence of terms that frequently co-occur in changes and failing tests. This empirical approach requires previous test failures and thus either bootstrapping, which we will achieve through mutation testing, or a documented history of changes and test runs, as found in continuous integration infrastructure.

**Context** By using only the change as reported by a Unix-diff-like algorithm, the approach remains largely language agnostic, but the lexical context within the program structure (e.g. the surrounding method or class) is lost. Often, tests call a method or instantiate a class under test, while the change only affects the implementation thereof. The lexical context, e.g., the class or method name in which a change occurred could reveal which test is affected. In this work, we will explore the influence of including context in the query.

**Synonymy and Semantic Relatedness** A TF-IDF model would only consider exact matches of features. Synonyms (e.g. *count / number*, *str / string*, ...) and semantically related words (e.g. *draw & color*, *file & read*, ...) are regarded as distinct. For example, a change in a *color*-related method may affect *drawing* tests, but their relation is not reflected yet. A common approach is to compress semantically related words into *topics* using a *topic model*, such as Latent Dirichlet Allocation (LDA) [2], but no studies managed to conclusively show improvement of prioritization performance, or whether measured improvements justify such a complex model [11]. Topic modeling is computationally expensive, requires fine-tuning, and can negate the time savings of prioritization.

**Abstractions** On closer consideration, semantic relatedness is often *asymmetric:* Drawing routines use colors, hence the *drawing* test may fail more likely when color-related code is modified, but no *color* test should fail when *drawing* routines are updated. One feature (*drawing*) is at a higher level of abstraction, and *color* is an implementation detail. While a call graph analysis or coverage data includes this aspect, lexical change data does not.

**Updatability** A simple TF-IDF-based model can be cheaply updated when a new test is introduced or an existing one is changed, as only IDF scores of the affected features need to be recomputed, as well as TF scores of the new/changed test. Any model involving topics, abstractions, or machine learning is likely much larger as it is concerned with connecting semantically related features beyond those found directly in tests. These models would need to be updated after any code change, or fully re-trained. Coverage-based models are effectively updated using test runs themselves, making them outdated the moment they are used in a predictive setting.

## 3 Fault Seeding

If historical code changes and subsequent testing reports from a program were available, we could evaluate test prioritization strategies without user involvement. Simulating the impact a “treated” test order would have had on the report and comparing metrics (e.g., position of first failure) helps to estimate how much faster the desired feedback would have been obtained if the treatment had been in place.

Unfortunately, realistic examples of changes causing test failures rarely leave traces in publicly available data. Most programmers ensure all tests are passing before committing their change to a public repository, and only in infrequent cases did a continuous integration (CI) server (e.g. TravisCI) produce a detailed log of a failed build. In light of this data scarcity, we see the need to synthesize faulty changes and corresponding test results.

A strategy used by the *REPiR* project [18] is focused on sampling *regressions* by running the regression test suite of an old version against a newer version of the program. This way, the old test suite lacks tests for new functionality, but the changes that lead to test failures represent real programming activity. In contrast, our approach aims at keeping test suite and program in sync, while relaxing the requirement that the fault be caused by the change.

First, we elaborate on our general fault seeding strategy and analyse an example run to highlight the characteristics of the test suites. Second, we adapt the fault seeding process to follow the actual distribution of program changes using historical revisions and generate the synthetic test failures that underlie our evaluation.

### 3.1 Mutation Testing

The practice of mutation testing was originally designed to judge the quality of a test suite: if the mutation tester could alter a syntactic expression without any failing test detecting this change, there is a high probability that no test logic covers the implemented logic that was being modified. We adopt this practice as a way to generate failing test runs as a result of small program changes (mutations).

We exclude non-code files (documentation, configuration, resources, build scripts, etc.). Moreover, test code is excluded from fault seeding, because compromising one test should not cause any test other than the faulty one to fail if best practices have been followed.

Our fault seeding tool is designed for and written in Python. The approach is general enough to be applied to other dynamic languages but is not guaranteed to generate a valid program under static type systems.

**Mutation Operators** Operators are the core components of mutation testers. In our case, an operator is an Abstract Syntax Tree (AST) transformation that consists of two rules: A definition of AST nodes to which it is applicable, and a transformation that takes an AST node and returns a new one replacing the originally matched node.

We make use of similar operators as used in the JAVALANCHE [19] tool. Their representativeness has been partially confirmed by literature [1]. A notable difference is that we are working on ASTs rather than bytecode for practical reasons: since the parser keeps track of the origin of each node, it is easier to link AST nodes with modified lines of code in a Git commit, which we will use in our change-based fault seeding approach. In contrast to Java infrastructure, our test runner operates on source directories rather than compiled artifacts, hence a source-to-source transformation makes sense.

**Negate Branch Condition** Applies to AST nodes in the role of an if-condition. Mutation returns the AST node wrapped into a unary negation node. The operator does not apply inside potentially infinite (`while True`) loops to prevent timeouts.

**Table 1** Examples illustrating applicability and effect of mutation operators

| Operator | Matched Code | Transformed Code |
|---|---|---|
| Negate Branch Condition | `if arg < 0:` | `if not (arg < 0):` |
| | `elif arg > 100:` | `elif not (arg > 100):` |
| Omit Method Call | `super().__init__(foo())` | `None` |
| | `super().__init__(foo())` | `super().__init__(None)` |
| Swap Arithmetic Operator | `size = end - start` | `size = end+start` |
| Modify Number | `count += 1` | `count += 2` |

**Omit Method Call** Applies to Call nodes. Mutation returns the `None` literal. If the call was a statement, it has no side effect anymore. If it was part of an expression, it simulates forgetting to return a value.[1] This operator is also forbidden within infinite loops.

**Swap Arithmetic Operator** Applies to AST nodes of binary operator type. Mutation swaps `+` with `-` and `*` with `/`. For simplicity, we ignore logical operators (their frequent use in conditions is already covered by Negate Branch Condition), rarely used bit-operations, and the modulo (`%`) operator which is primarily used for string formatting in Python.

**Modify Number** Applies to AST leaves representing numeric literals. Mutation increments them by 1 with the intent of causing off-by-one or indexing errors. However, many numbers have non-functional roles, such as buffer sizes and timeouts, or are exchangeable by design, such as port numbers or error codes.

In a first step, each operator is applied to the program's AST and all suitable locations in the code are collected, no modification happens yet. The result is a set of *candidate locations*. For each of these candidates, the unmodified AST is copied in memory, the chosen candidate location modified, serialized, and written to a working directory in which the test runner starts the (unmodified) test suite.

**Distribution and Impact** As an example, we run all mutations on the *Flask* web framework, one of the most popular Python software packages that is tested with an unmodified *PyTest* test runner. In total, 1249 candidate locations have been identified, the percentage of each operator is shown in figure 1a, omitting calls making up the majority. 1059 of them caused at least one new test failure besides the control run on an unmodified code base. Failure causes can be seen in figure 1b.

The failure rate in test runs follows a multimodal distribution, which can be seen in figure 2a. We observe two clusters of seeded defects: The left cluster contains "low-impact" defects that trigger a few (< 100) failing tests, the right cluster of "high-impact" defects caused half or more of the test suite to fail. This can be explained by the degree to which the defect is isolated: Defects inside individual methods are usually caught by their specific tests, while defects at module level (executed at import-

[1] In Python, exiting a function without return results in `None`.

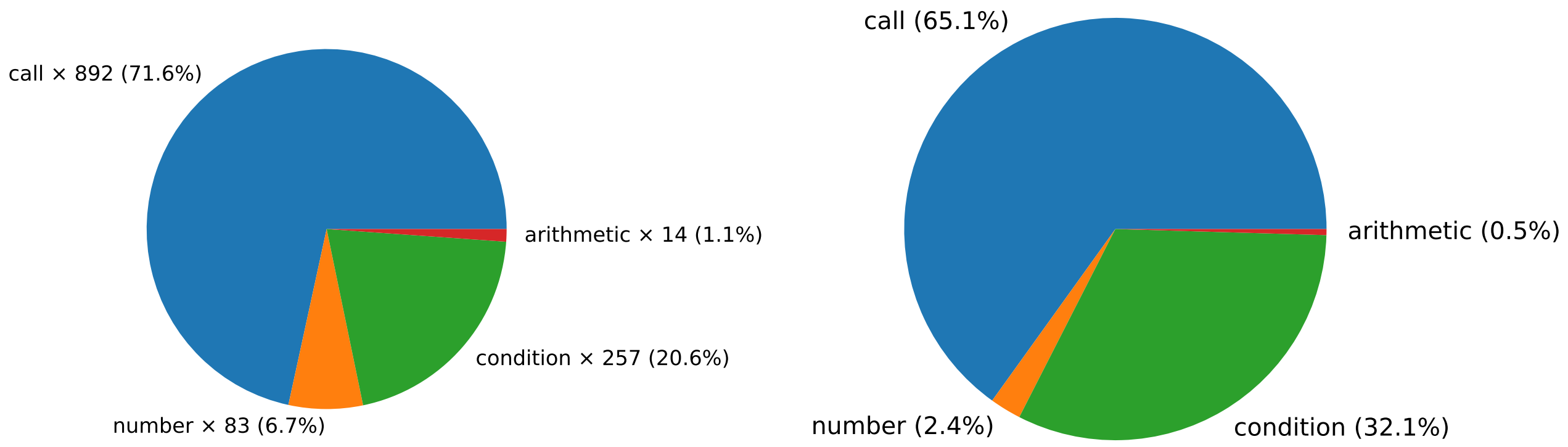

**(a)** Operators of seeded defects

**(b)** Causes of test failures

**Figure 1** Distributions of the number of failing tests in the Flask project in each run (left) and the number of seeded defects detected by each test (right).

time in Python) can result in improperly initialized components with widespread consequences. The most prominent reason for failing tests is `AttributeError`, primarily caused by accessing a field or calling a method on `None`, i.e., the analog to a null pointer exception.

The test suite itself has mostly specific tests that only detect a small part of the seeded defects, with another cluster of tests that detect about one tenth of the seeded defects, as shown in figure 2b. This split in specific (left) and sensitive (right) tests can be explained by the difference between "canonic" *unit* tests and *integration tests*: Canonic unit tests isolate the unit under test and generally run only a small fraction of production code, while integration tests check whether a composition of multiple components cooperates as specified. The integration tests in the right cluster of histogram figure 2b test complex framework activities, like file transfers, security-related aspects, and request routing in the presence of errors.

### 3.2 Change-based Fault Seeding

To obtain synthetic, yet realistic, faulty changes and associated test results, we propose a fault seeding strategy based on the actual edit history of the program. We compare each version (Git commit) to the previous version and only changed or inserted lines are considered for seeding faults, i.e., our candidate locations are intersected with that line-based diff. The distribution of faults still resembles that of figure 1 but is distributed like actual code changes.

If the candidate set is empty, this particular version is uninteresting and we proceed to the previous version (i.e., the parent Git commit). This usually happens when documentation or configuration fixes are being committed, but no change in program logic.

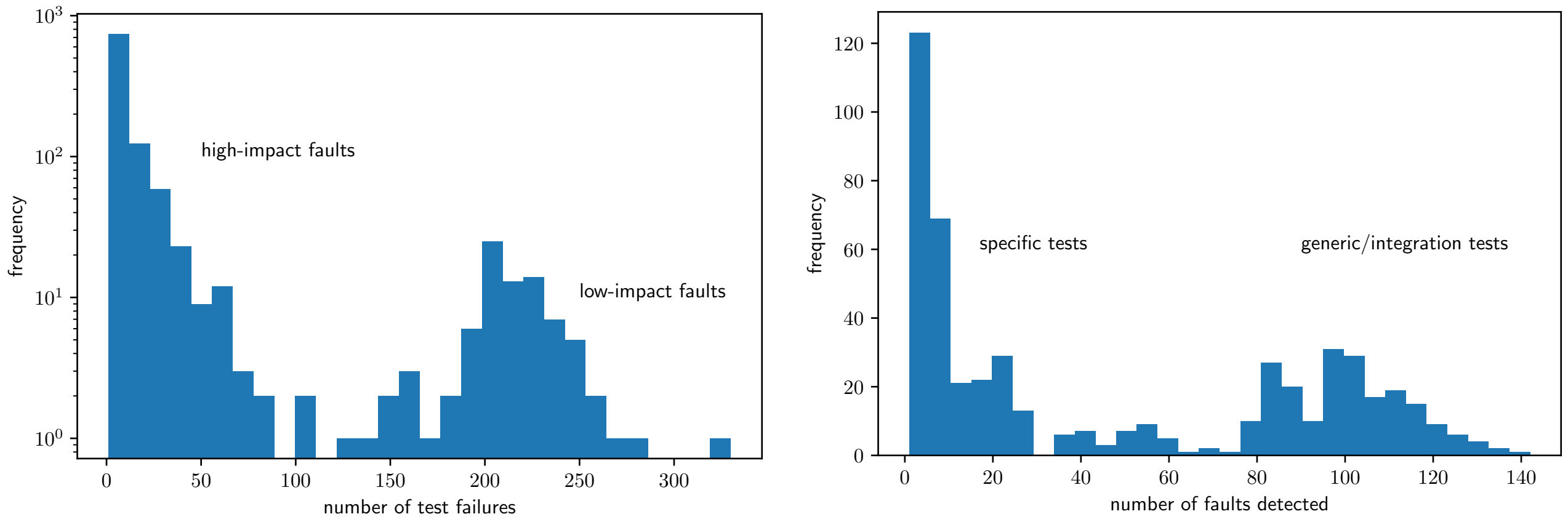

**(a)** Histogram over the number of failing tests per test run/seeded defect. Runs on the left contain more succeeding tests, tuns on the right more failing tests.

**(b)** Histogram over the number of faults detected per test. Tests on the left were highly insensitive, while these on the right failed more frequently.

**Figure 2** Distributions of the number of failing tests and the number of seeded defects detected by each test in the Flask project.

Previous work successfully used a similar approach in Java to prioritize tests based on the rate of mutants detected by each test [9].

**Collecting Test Results** The collection process starts with the most recent version of a software repository and then iterates backwards through the main line of changes. This way, we can use an up-to-date test framework that works with recent versions and stop the iteration once a version is old enough to be incompatible to the current set of dependencies.

For each version, we first do a *control run* of the unmodified test suite with the unmodified program. For each candidate, we collect a *single-fault result* by running the test suite with the program that has exactly this single mutation applied. For further processing of single-fault results we only consider tests that passed in the control run and were executed in the presence of the fault, too.

## 4 Relevance Scoring

The IDF weighting of individual terms in the TF-IDF approach does not reflect the importance of a word with respect to predicting failing tests. With current test runners and Continuous Integration (CI) infrastructure, however, it should be possible to directly collect evidence of the *predictive power* of any term or lexical feature as defined in section 2.

**Sampling Predictive Power** If a feature appears in a *faulty* change, then this feature is said to predict a test failure accurately when the following statistical criteria are met:

1. From all tests that contain this feature, most are failing in response to the faulty change. If we prioritize tests based on this feature, this criterion avoids ranking passing tests too early (minimizing false positives).
2. The set of tests containing this feature make up a large portion of the overall failing tests. This prevents prioritization based on this feature from missing failing tests and letting them move to the lowest ranked positions (minimizing false negatives).

We can express both proportions in terms of (1.) *precision* ($PREC$), (2.) *recall* ($REC$), and the $F_1$*-measure*, which is defined as harmonic mean of precision and recall and the conventional way to combine both quantities [14].

Let $F$ be the number of failing tests, $F_w$ the number of failing tests containing feature $w$, and $T_w$ the number of (either failing or passing) tests containing $w$.

$$PREC(w) = \frac{F_w}{T_w} \tag{2}$$

$$REC(w) = \frac{F_w}{F} \tag{3}$$

$$F_1(w) = 2 \cdot \frac{PREC(w) \cdot REC(w)}{PREC(w) + REC(w)} \tag{4}$$

If a feature does not co-occur in change and test, it should have zero weight. We combine this binary term frequency (0 for an absent feature, 1 for a co-occurring feature) with the three quantities to obtain Term Frequency–Precision (TF-PREC), Term Frequency–Recall (TF-REC), and Term Frequency–$F_1$ (TF-$F_1$) as alternatives to TF-IDF.

The resulting score function for a change vector $c$ and a test vector $t$ becomes:

$$S(c,t) = \sum_{w \in c} P(w) \cdot TF(w,t) \tag{5}$$

with the predictive power weight $P$ being either $PREC$, $REC$, or $F_1$ of the feature or word $w$ and $TF$ being the binary term frequency that effectively drops the complex document length correction from BM25.

To prepare our study, we collect data points for each word $w$ during mutation test runs and average precision, recall, and $F_1$ score over all test runs, excluding those not containing $w$ in a change.

**Window** By including the neighbourhood of a change, i.e., a window of $\pm n$ lines of code around a seeded fault location, the average accuracy of each feature can be improved up to $n = 2$ (we sampled scores from $n = 0$, i.e. only the line a change occurred in, up to $n = 16$), which indicates up to 5 logical lines of code surrounding a particular mutation contain lexical features of predictive power. We use windowing in an attempt to mitigate the limitation of sampling the scores from mutation testing rather than real changes, e.g., in a development environment that can keep track of fine-grained changes and the resulting unit test runs.

**Weight distribution** We observe that precision is biased towards rare words, since they have little chance to co-occur in a passing test and a change, while recall is biased

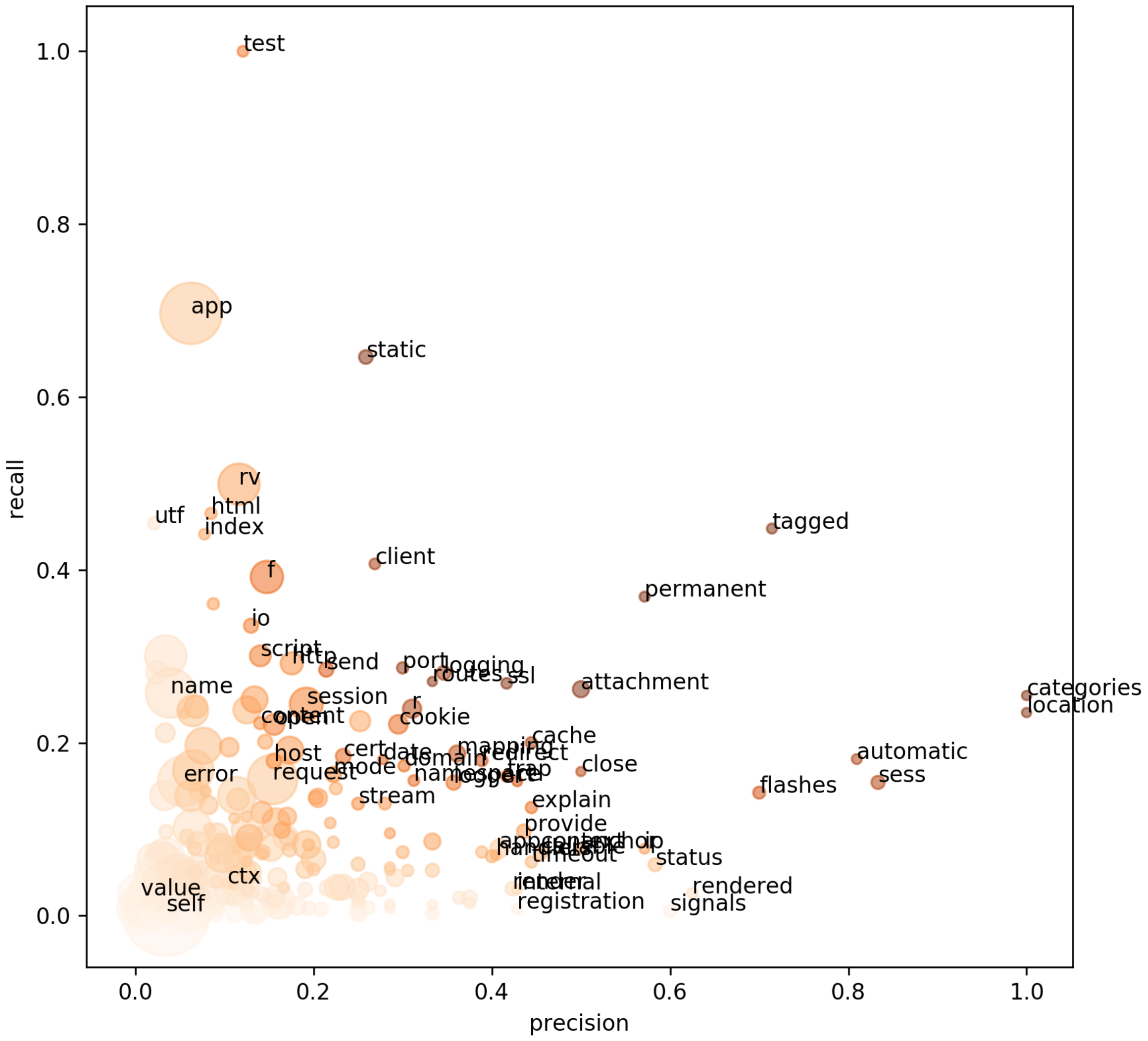

■ **Figure 3** Predictive power of individual lexical features in Flask. The x-axis denotes the probability of a test failing when it contains the word (precision), the y-axis the percentage of failing tests identified by the word (recall). The size of each bubble is proportional to the word's overall frequency, darker orange tones indicate higher $F_1$-scores.

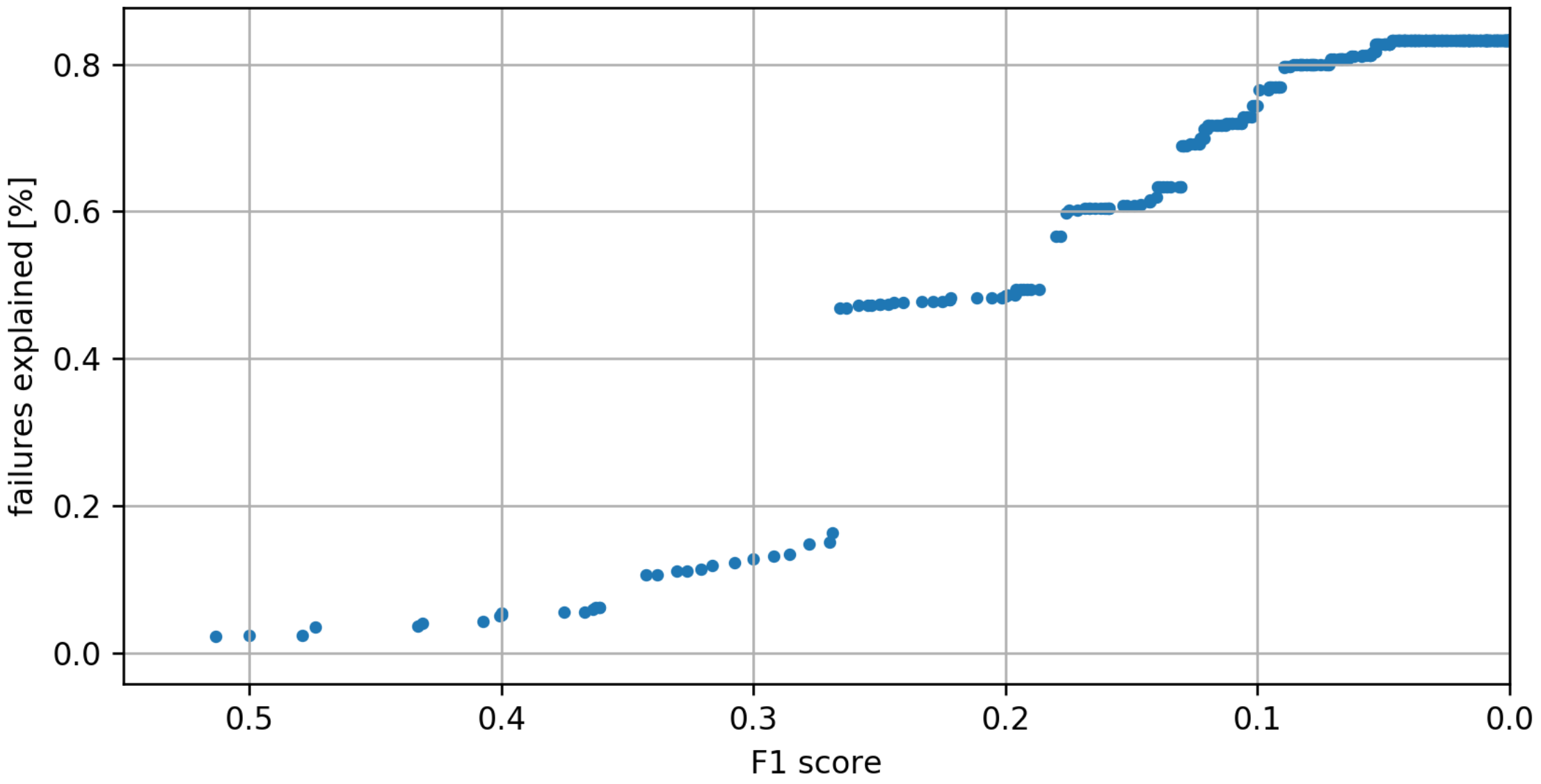

**Figure 4** Cumulative percentage of test failures explained by words with the top $F_1$ scores. The median is at $f_1 = 0.18$ after the top 62 words.

towards words that occur in many, and thus most failing tests. In figure 3, precision, recall, and F1 scores of each identifier in the Flask project can be seen. The higher $F_1$ scores towards the top right corner belong to specific concepts of the web framework, for example static files, sessions, SSL, logging, or caching. As expected, meaningless words like `self`[2] rank low in both dimensions.

With respect to the whole test suite, the words with highest predictive power help to explain only a small percentage of test failures. In figure 4, we sorted words by F1 score (most predictive words on the left) and accumulated the percentage of test failures that would be predicted by these words so far. The graph shows that, in order to explain half of the test failures, we need to consider all 62 words with $F_1 > 0.180$. Even when considering all lexical features, which would introduce a high number of false positives, this approach is limited to explaining 83 % of failures in Flask. This shows a general limitation of using lexical features, as the remaining 17 % of tests failed without overlapping vocabulary.

## 5 Experiment

To assess the effectiveness of different variations of lexical test prioritization, we ran an experiment on four GitHub projects written in Python. We investigate the following prioritization strategies:

[2] Which is not a keyword in Python but analog to this in Java.

**UNT** The native (untreated) order of the test suites as chosen by the test runner.

**RAND** The tests in random order.

**BM25** Tests ranked according to their relevance score assigned by the BM25 model with $k_1 = 10$ and $b = 0.5$.

**BM25C** The same as BM25, except the change query includes lexical context, i.e., class and function names in which the change occurred.

**PREC** Tests ranked according to their TF-PREC score

**REC** Tests ranked according to their TF-REC score

**F1** Tests ranked according to their TF-$F_1$ score

Our setup assesses the following questions:

1. Does IR-based test prioritization (BM25) improve test ordering over the native ordering (UNT) and random ordering (RAND) in Python?
2. Does inclusion of contextual features (BM25C) improve prioritization in TF-IDF (BM25)?
3. What is the impact on test ranking when weighting by predictive power (PREC, REC, or F1) rather than IDF in TF-IDF (BM25)?

**Project selection** With Python, we have chosen an interpreted language that avoids build overhead before test runs and between different program versions. The *PyTest* test runner is easy to instrument. We require our projects to be testable with PyTest directly from their download location given their dependencies are installed, and belong to the top 100 in popularity on GitHub (measured by number of stars after selecting those marked as Python) at the time of writing. Unexpectedly, almost none of the projects on GitHub turned out to be testable out of the box without compiling separate libraries, requiring a modified test runner, or a specific platform to run the tests. This left us with four popular projects to sample from[3]:

**Flask** After Django (which uses its own test runner) the second most popular Python web framework

**Requests** Client library for web requests

**Sphinx** Documentation generator

**Jinja** Template language and templating library primarily used for HTML responses in web servers

**Fault seeding** For each project and each Git commit, we perform a control run of the test suite (usually located in the tests directory) with Python's import path set to the checked out working tree of that commit. If the control run was able to report at least one successful test, all *.py files that have changed since the previous Git commit are gathered (via the *GitPython* library) and the differing ranges of lines are computed using Python's `SequenceMatcher`.

[3] The PyTest project itself fulfills these requirements but had to be dropped. Our setup could not sufficiently isolate the colliding namespaces of the PyTest instance we use for testing and the faulty PyTest version being tested.

**Table 2** Number of commits, total number of faults seeded (= number of test runs collected), average number of tests executed per run, and failures per run.

| Project | Commits | Seeded faults | Tests/run | Failures/run | Failure density [%] |
|---|---|---|---|---|---|
| Flask | 74 | 451 | 400.0 | 13.0 | 3.2 |
| Requests | 121 | 469 | 72.8 | 2.4 | 3.3 |
| Sphinx | 30 | 410 | 1145.2 | 182.4 | 15.9 |
| Jinja | 59 | 496 | 436.0 | 24.7 | 5.7 |

Subsequently, the files are parsed using the `ast` module to obtain their ASTs. Mutation operators, implemented as `NodeTransformer` subclasses (a mutating visitor pattern), are then applied in a "dry run" to gather candidate mutations. Candidates are identified by a mutant ID (concatenation of file path, operator name, line number, and column offset) and discarded when their line is not within one of the previously computed diff ranges.

For each remaining candidate, the operator that generated it is applied to the exact location only, the resulting ASTs are written back into the Git working tree, and PyTest executed in a separate process. PyTest is instrumented using a plugin that writes (test, success, duration)-tuples to a file that is identified by commit hash and mutant ID. Individual tests that did not succeed in the control run, and full test suite runs that match their control run exactly are discarded, since they did not detect failures caused by the seeded fault.

Flask and Requests began to show incompatibilities to the installed dependencies (we opted not to re-install dependencies for each version, since this would increase sampling time by orders of magnitude) and the sampling process for Sphinx exceeded 48 hours. Faulty test runs were interrupted after 10 minutes (approximately double the time of the longest non-faulty *control run*) and discarded from analysis in case the seeded defect prevents termination.

The number of selected commits, successfully seeded faults (which is equivalent to the number of obtained test suite runs), and the average number of tests per run is listed in table 2.

**Test indexing** For each test run, we parse the source files containing the executed tests and extract the following features using a `NodeVisitor` subclass:

- Name of the test method
- Names of fixtures and parameters (in parametrized testing) given to the test
- The test method's documentation ("docstring")
- Identifiers (ASTs nodes of type `Name`)
- Strings

Composite features are broken down and normalized to lower case, e. g. `CamelCase` results in `camel` and `case`, `HTTPServer` becomes `http` and `server`, and any non-alphabetic character is considered a delimiter.

**Change query construction** The query against the set of tests is constructed analogously to the feature extraction described for tests. As discussed above, the context of a change is not reflected in line-based diffs but might be important. To be able to assess the value of the outer lexical context, we provide two ways to construct a query from a change:

**Without Context** Used in BM25, REC, PREC, F1. The visitor, although it scans the whole ASTS, only emits features when the ASTS nodes lie within the diff to the previous commit.

**With Context** Used in BM25C. The visitor carries class and function names along and adds them to the list of features when a change affects the respective class/function.

### 5.1 Measurements

To quantify the effectiveness of a prioritization strategy, we employ the Average Percentage of Faults Detected (APFD) metric. We then measure the average time to detect the seeded failure over all test suite runs. A graphical comparison is given by plotting cumulative detected faults over run-time of the test suite.

**APFD** The APFD metric is frequently used to assess the effectiveness of a prioritized test suite [8]. It is defined over a set of faults $F = \{f_1 ... f_m\}$ and a sequence of tests $t_1 ... t_n$ as:

$$\text{APFD}(F; t_1 ... t_n) = 1 - \frac{\sum_{f \in F} TTF(f)}{n \cdot m} + \frac{1}{2 \cdot n} \tag{6}$$

where $TTF(f)$ represents the position of the first test that exposes fault $f$. When reporting APFD, we average values over all test runs. When plotting the proportion of failing tests over the proportion of tests executed so far, this metric measures the area under this curve, with higher numbers being better, and 50 % corresponding to equally spaced test failures across the run.[4]

In table 3, APFD scores (in percent) for each evaluated prioritization strategies are given. Averages hint at a slight advantage of precision-based ranking over standard BM25. Including contextual data is no improvement over normal BM25 ranking with respect to the APFD metric.

**Significance** To assess if the effects are significant, we perform a Wilcoxon signed-rank test. This is a *paired* test, as APFD is measured on the *same* test suites ranked according to the different strategies. We reject the null hypothesis that a prioritization

[4] The metric's range is the open interval $(0, 1)$ where 0.0 or 1.0 are approached if the number of passing tests before or after the failures approach infinity.

**Table 3** Mean APFD scores (averaged over all seeded defects) in *percent* with standard deviation. Higher values are better.

| Project | UNT | RAND | BM25 | BM25C | REC | PREC | F1 |
|---|---|---|---|---|---|---|---|
| Flask | 43.8 ± 11.6 | 50.8 ± 13.1 | 65.5 ± 13.6 | 65.4 ± 13.1 | 67.1 ± 15.7 | **71.7** ± 13.7 | 70.9 ± 13.8 |
| Requests | 63.8 ± 14.3 | 51.0 ± 9.5 | 69.1 ± 10.8 | 68.5 ± 11.0 | **75.7** ± 9.8 | 74.6 ± 9.1 | 75.4 ± 9.2 |
| Sphinx | 48.9 ± 13.2 | 49.8 ± 4.0 | 58.6 ± 13.0 | 59.0 ± 13.7 | 57.9 ± 13.6 | **60.9** ± 13.5 | 60.5 ± 13.6 |
| Jinja | 46.3 ± 23.7 | 48.6 ± 16.0 | 82.9 ± 18.2 | 82.6 ± 18.7 | 81.7 ± 16.7 | **84.4** ± 17.1 | 83.8 ± 16.3 |

was not effective over a baseline from $p < 0.001$. We also compute the percentage of test runs that improved over the respective baselines:

**BM25 over UNT** The general IR approach is *not* significantly better when applied to *Requests*, but it outperforms the untreated baseline in the other three projects. Over all projects, BM25 ranked 90 %, 59 %, 60 %, and 81 % of test runs better than the native ordering.

**BM25 over RAND** The general IR approach is always significantly better than random. BM25 ranked 79 %, 92 %, 62 %, and 91 % of test runs better than random.

**BM25C over BM25** Using context information does *not* significantly improve prioritization over a general IR approach.

**PREC over BM25 and UNT** Precision-based ranking (TF-PREC) appears to be the best of the measured strategies, being *significantly* better than all TF-IDF-based and native test orders. PREC ranked 77 %, 79 %, 68 % and 51 % of test runs better than BM25. Whenever BM25 improved a test run over UNT, PREC could further improve the schedule in 60 % of the cases. This suggests that test runs which are "prioritizable" using TF-IDF continue to improve with precision-based ranking.

REC and F1 are not discussed further, as they failed at least one significance test.

**Execution time** While APFD scores are independent of time measurements and reproducibly comparable over test suites, a second construct we are trying to optimize is expressed in time to detect a defect. We measured the execution time of each test and accumulate the time spent in unit tests until the defect is first exposed. Plotting the amount of defects detected up to a point in time gives a profile of how effective the prioritized order is over time. Both are illustrated in figure 6.

We assume the durations of individual tests are insensitive to their ordering, hence we are re-using durations measured during the untreated test run.[5] The highly I/O-sensitive set-up time needed to discover and load tests from the file system, as well fixture initialization and the progress output to console are not included.

[5] Benchmark system: Commercial off-the-shelf laptop with Intel Core i7-8650 CPU (4.2 GHz), 16 GB memory, 512 GB SSD, running Windows 10 17763 in developer mode with disabled malware protection. Instrumented test runner was PyTest 4.6.5 on Python 3.7.4 x64.

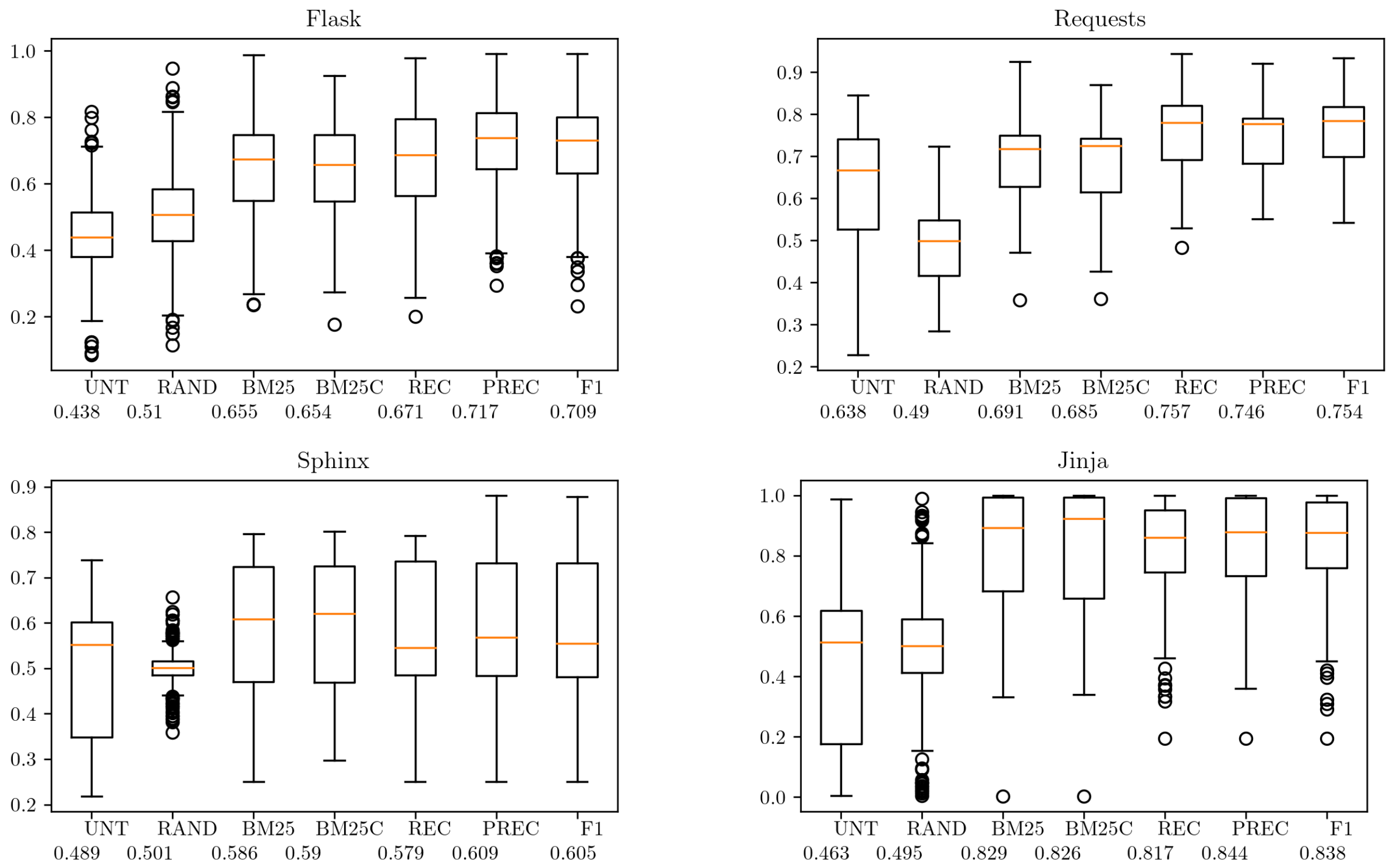

**Figure 5** APFD scores of Flask, Requests, Sphinx, and Jinja for each prioritization strategy.

Measurements show strong variability (outliers omitted). Although average times to detect the fault are consistently and significantly[6] lower with BM25 when compared to untreated and random baselines, the benefits of accuracy-based weighting schemes (REC, PREC, F1) over traditional IR become less pronounced with respect to average timing. Consistent with our APFD findings, REC and F1 are not beneficial. Precision-based weighting is statistically indistinguishable from BM25 regarding timings.

The measurement of average time to detect the first fault is complemented by the curves that show a steep increase in detected faults in the very beginning of the test suite run for every lexical prioritization strategy. In the largest project, Sphinx, we even observe an advantage of precision-based weighting over BM25, with the remaining graphs showing high similarity of all profiles.

**Discussion** Our experiments, although only taken from four well-tested Python projects, are consistent with related work on regression testing and provide additional evidence that lexical test prioritization is able to shorten fault detection times up to a level suited for immediate feedback.

We are dealing with a dynamic language that lacks the verbosity of static types. Consequentially, our prioritization has access to fewer lexical features. Related work on Java projects [18] achieves similar levels of APFD scores using TF-IDF (*REPiR* on method level: 64 % to 87 %), and our results provide evidence that TF-IDF prioritization works almost as effectively in Python as confirmed by APFD scores and time to fault detection.

[6] $p < 0.001$ via Wilcoxon signed-rank test.

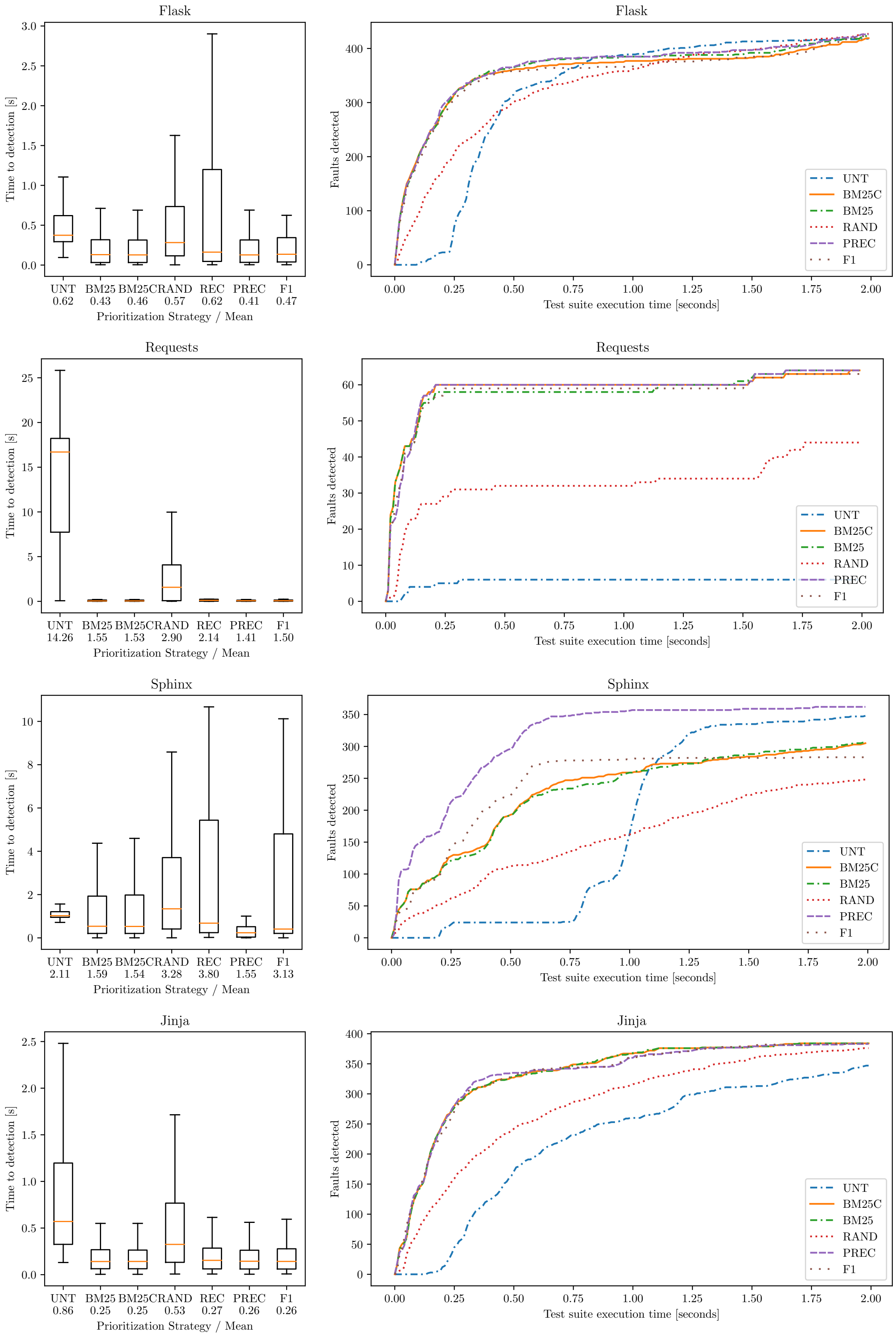

**Figure 6** Time to detect a fault and cumulative faults detected over time for the example projects

The high variability in measurements is likely a combination of several factors; the observation from section 4 suggests that individual tests can fail without overlapping vocabulary, while manual inspection revealed that the worst prioritization results were caused by very large changes with many lexical features. Both change and test characteristics can determine "prioritizability" of a test suite run.

When weighting terms using predictive power over classical IR schemes, we found that using *precision* over IDF significantly improves how early failing tests are scheduled with respect to their position in the test suite (as measured by APFD), and can also safely conclude that TF-PREC outperforms untreated and random ordering.

However, weighting lexical features by *recall* or *F1 score* did not produce significant improvements. None of our newly proposed strategies outperform TF-IDF with respect to the time it takes to find the first failure, but they are not worse either.

### 5.2 Limitations and Threats to Validity

**Large Changes** A GitHub repository tends to accumulate larger changes when contributions from other users are accepted. These so called *pull requests* are often multiple changes folded into what appears as a single change in our traversal of the main line of development and thus introduce noise.

**Preprocessing** Because many concepts in a program use compound names, our preprocessing step splits camel case to recognize the individual subconcepts, e.g. a change to a `session_cache` variable or the `SecureCookieSession` class would be connected with any test using a `session` fixture. While inspecting examples where test prioritization failed, we realized that this construction can as well have the drawback of decomposing a meaningful identifier into meaningless parts. An example is the `NoAppException` class, a specific concept that would co-occur in some exception handlers and tests, but the constituents `no`, `app`, and `exception` are very generic and receive low IDF and accuracy weights.

**Time Accuracy** Some unit tests in our example projects can run in under a millisecond, while some are in the order of seconds. We explicitly used the newest version of Python at the time of writing, which provides a nanosecond-resolution timer, but given operating system constraints, the inexactness of the timestamp can be within the order of magnitude of a fast unit test. Also, we did not measure overhead introduced by, e.g., fixture creation. However, we expect many of these effects to cancel out statistically over the number of data points that have been sampled.

**Synthetic Defects** The use of mutation testing to approximate programmer-induced defects is controversial. An early study by Andrews, Briand, and Labiche found that they sufficiently predict the real-world fault detection capability of tests [1], while Just, Jalali, Inozemtseva, Ernst, Holmes, and Fraser identified numerous real-world faults that cause failure modes which frequently used mutation operators cannot re-produce [5]. Common real-world defect types that are not represented by our strategy include mis-placed break/continue statements, confused argument order (especially of arguments with the same type), using similar but wrong method/function names, and accessing fields when a non-trivial getter/setter should have been used. Most of these defect types have been analyzed in Java with limited generalizability for dynamic languages.

**Generalizability** The scope of our conclusions is limited given the small selection of projects in a single language. Moreover, they are well tested compared to projects with fewer users and contributors, and mostly relate to web-centered topics where the Python ecosystem sees widespread use. The size of the test suites is less than 2000 tests (including parametrized tests) and they run in less than 10 minutes on consumer hardware, so we cannot generalize this to larger test suites yet.

## 6 Future Work

**Differential Retrieval** Ranking tests according to their lexical similarity to a change causes the top-ranked tests to be very similar. Having a more diverse set of concepts covered up-front could increase the usefulness of the information conveyed early during the test run. Instead of ranking all tests using the same score, tests could be selected one-by-one, after each the score would be re-adjusted to select the test that best covers the *remaining* vocabulary not yet matched by the already selected tests. We hypothesize that a *differential* ranking could prove powerful for larger changes that modify more than one concept.

**Approximate Coverage** Currently, we have no representation for the fact that some concept (e.g. a UI element) would eventually end up using a set of different concepts (e.g. drawing routines for displaying the element, event handling for processing user interaction), in which case the vocabulary of the implementation-specific concepts would not help prioritizing a test for the (dependent) UI element. Using lexical information alone to model such a call-graph is an interesting challenge and could be realized by keeping track of how often each word from a definition (e.g. method name) leads to another word from its implementation (e.g. a called method). This Markov chain can be used to estimate the probability that any test *transitively* depends on the changed vocabulary.

**Increasing Predictability** A drawback of change-based information retrieval is that each time the program changes, programmers will be surprised by a completely new test order. We see two directions for solving this problem:

1. Improve tooling to identify relevant tests while the change is being made, so that programmers have the chance to take note of the new test ordering, or even intervene, before tests are executed.

2. Stabilize the (re-)ordering of tests by taking previous orders into account and keep the change visually small, e.g. an important test is inserted at the front, while the order of less relevant tests is kept unchanged.

**AutoTDD** Our proposed tool integration targets live programming environments, such as *Squeak/Smalltalk*.[7] We propose a workflow in which tests are executed after every modification, i.e., saving (via ctrl + s) in Squeak. Existing testing frameworks, such as *SUnit* in connection with edit-triggered test runners (*AutoTDD*[8]) already provide this functionality. They can be extended to collect edit features and prioritize tests accordingly. Additional priority can be given to recently edited tests and recently failed tests.

**Continuous testing** A yet unexplored challenge emerges when test runs are restarted after every edit, leaving not enough time for low-priority tests. A *continuous testing* approach must balance the probability of catching an error against the uncertainty associated with not running a test for a long time. Relevant insights could be gained by studying the granularity and time intervals in which changes are saved in a live programming environment, how these relate to test execution times, and how much of an asynchronous delay programmers are willing to accept to maintain the impression of causality and continuity.

**Live examples** A novel approach to integrate example data with source code has been designed by Rauch, Rein, Ramson, Lincke, and Hirschfeld [12]. Especially in live programming environments, tests and examples serve a similar role in providing immediate feedback. Both are conceptually interchangeable, i.e., (failing) tests can provide explorable examples while an example with an assertion/expectation becomes a test. If a large set of examples is present, the same prioritization strategies that help selecting important tests may help selecting live examples as well. Their goal could be to provide the user with examples that are most relevant in the context of the current editing task.

## 7 Related Work

**Lexical Test Prioritization** Saha, Zhang, Khurshid, and Perry studied the effectiveness of IR-based regression test prioritization with their *REPiR* project [18]. With a focus on regression testing, the authors generate test failures by running new versions of Java projects against the test suite of the previous version, thereby obtaining regression faults. Their approach uses the Okapi TF and IDF weights on both query and document features. A notable extension of classical IR is their strategy to classify features according to their role in the source code and to the type of change they appear in, which allows comparing different roles (e.g. added/deleted fields, added/deleted methods) separately. This approach has been successfully used for bug localization in the *BLUiR* system before [18].

[7] https://squeak.org/, retrieved 2019-01-30.
[8] https://github.com/hpi-swa-teaching/AutoTDD, retrieved 2019-01-30.

*Static* test prioritization operates without change data and tries to spread similar tests to cover many concepts (diversity) or most of the program (coverage) up-front [10]. Topic models have been used by Thomas, Hemmati, Hassan, and Blostein [21] to model these concepts as topics over lexical features and hence address the concept coverage problem from a lexical viewpoint.

**Other heuristics** Using lexical information is a rather novel heuristic to judge the relevance or fault-detection capability of a test with respect to a change or full program. Much more literature is concerned with structural or empirically observed properties of the program under test or the tests themselves. Although common, not all strategies fix a schedule beforehand but use just-in-time prioritization with access to results produced during the test run under consideration.

Important prioritization strategies are:

**Coverage-maximizing heuristics** Maximizing the code reached by the test suite as early as possible is the most common strategy. *Jupta* [22] maximizes methods covered based on a static call graph. The underlying technique was later extended with dynamic analysis to address limitations of static analysis in the presence of metaprogramming [20] for the purpose of regression test selection. Early coverage-based approaches focus on maximizing statement coverage and branch coverage, either by ranking tests based on absolute statement/branch coverage, or by first selecting the test with most coverage and then adding tests based on how many additional statements or banches they cover [16]. These tactics apply to change-based prioritization by intersecting coverage information with the set of changed statements, branches, methods, or components.

**Fault-oriented heuristics** Instead of using coverage as a proxy, some techniques measure fault-detection capability of a test more directly. Rothermel et al. used *mutation scores* for each test [16], which they considered much more expensive but slightly more accurate than a coverage-based heuristic. The complexity of functions covered by a test can be used to measure risk or error-proneness. This has been extended to change-based scenarios where the complexity of a change drives the ranking of tests, i.e., tests that cover the most error-prone changes are run first [4].

**History-based heuristics** With the presence of historical data about test runs, correlations between co-failing tests have been used by Kim, Jeong, and Lee [6]: When during execution a new test fails and its failure is historically correlated with another not-yet executed test, the other test can be run earlier. In contrast to coverage-maximizing and fault-oriented techniques, this approach cannot fix a test schedule beforehand, but has access to results of the current test run to improve schedules just in time.

A more direct approach has been taken by Li, Harman, and Hierons, in which they considered prioritization as a search problem that directly optimizes the target metric (e.g. APFD). They used greedy algorithms, *hill climbing* and *genetic algorithms* [7].

The cost of each of these meta-heuristics compared to the heuristics above are not considered.

**Different evaluation metrics** While APFD is the de-facto standard for evaluating test schedules, $APFD_c$ [3] is a generalization that can capture cost and fault severity. When using execution time as cost function, the fact that running multiple cheap tests can have a higher combined fault-detection potential than a single expensive test in the same time can be considered. Measuring the cost of an individual test is difficult in modern frameworks, as the test order affects when certain test resources are initialized, which can make tests using the same resource cheaper, but takes time up-front.

The *average percentage of X coverage* ($APXC$) does not consider test failures, but the rate of coverage. Instead of counting and weighting the tests until a fault was detected, the tests until a unit of type $X$ was covered are counted. $X$ can refer to statements ($APSC$), blocks ($APBC$), or branches/decisions ($APDC$) [7].

## 8 Conclusion

Unit testing and immediate feedback are not mutually exclusive as long as running the test suite yields the most relevant feedback instantly. We explored the area of lexical test prioritization and conducted a study that demonstrates how unit test suites of several hundreds of tests can detect defects faster when prioritized by the amount of vocabulary they share with the change containing the fault.

Our experiments with synthetic defects confirm earlier Java-based results in a dynamically typed language, where most strategies relying on static analysis are not applicable. Building on this, we evaluated novel strategies to weight words according to their empirically observed predictive power and found small, but significant improvements in the precision-based strategy over previously used information retrieval techniques. The results suggest that simple models, which are likely the least expensive in terms of implementation and run-time costs, already offer competitive performance.

Motivated by these results, we look forward to integrating test prioritization in live programming environments and studying the interplay between short edit cycles, feedback generated by unit testing, and the capability to interact with live objects and examples during programming activities.

**Acknowledgements** We thank Falco Dürsch, who through his work on history-based test prioritization contributed to our understanding of test prioritization in general; the HPI Research School for Service-Oriented Systems Engineering for supporting this research; Stefan Ramson and Patrick Rein for comments on the approach; and Tobias Pape for editorial and typographic advice.

# A Auxiliary Diagrams

### A.1 Test Suite Characteristics of All Projects

We present the response of all test suites to mutations as done in section 3.1. The results in figure 7 show the distribution of mutation impact (left) and test sensitivity (right). Regarding test sensitivity, both Flask and Requests show the distinction earlier described as *unit* vs. *integration* tests, while Jinja has a more even spread with most tests being highly sensitive to mutations. The Sphinx project has mostly insensitive and specific tests.

In figure 8, figure 9, and figure 10 we illustrate the predictive power of individual lexical features of the remaining three projects.

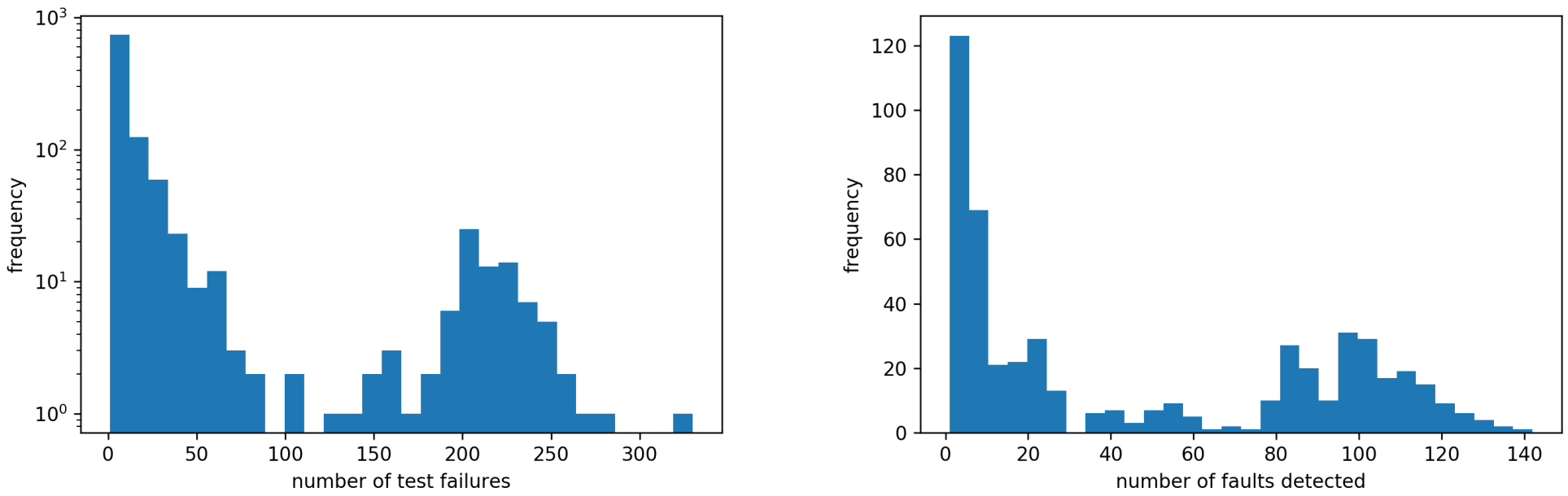

**(a)** Flask

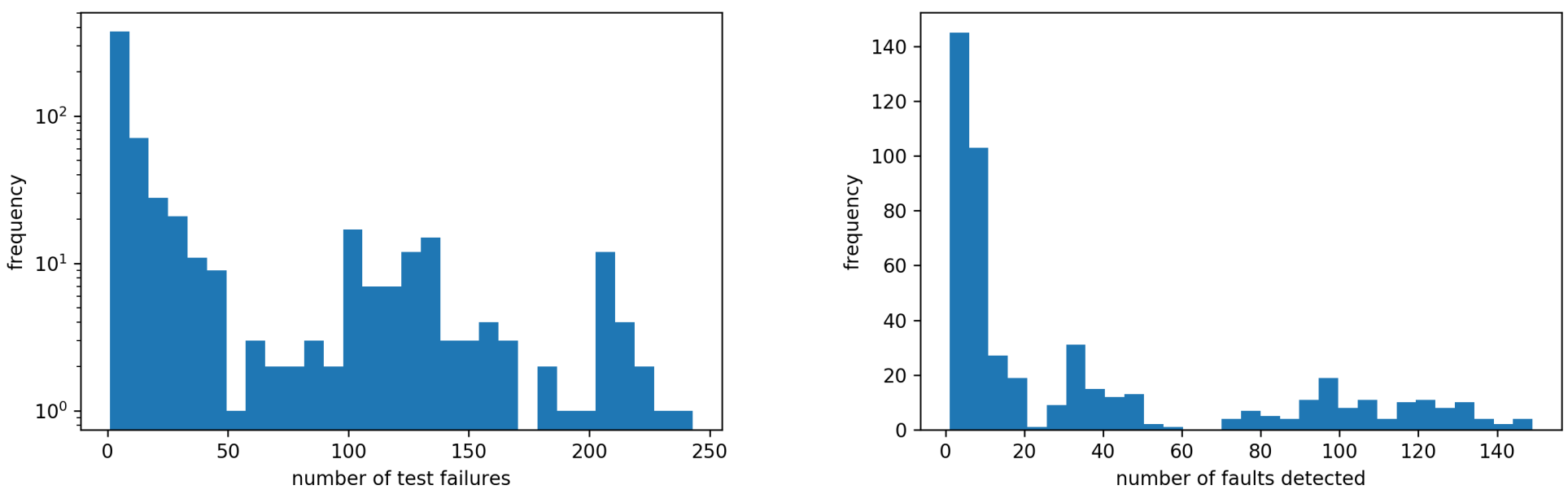

**(b)** Requests

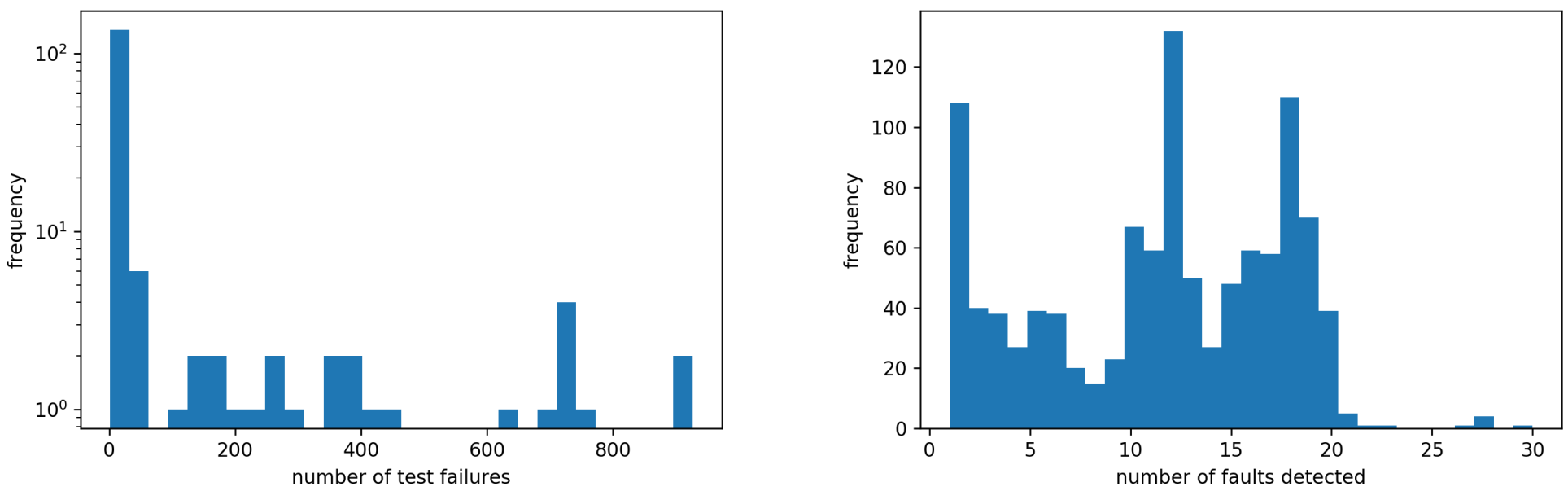

**(c)** Sphinx

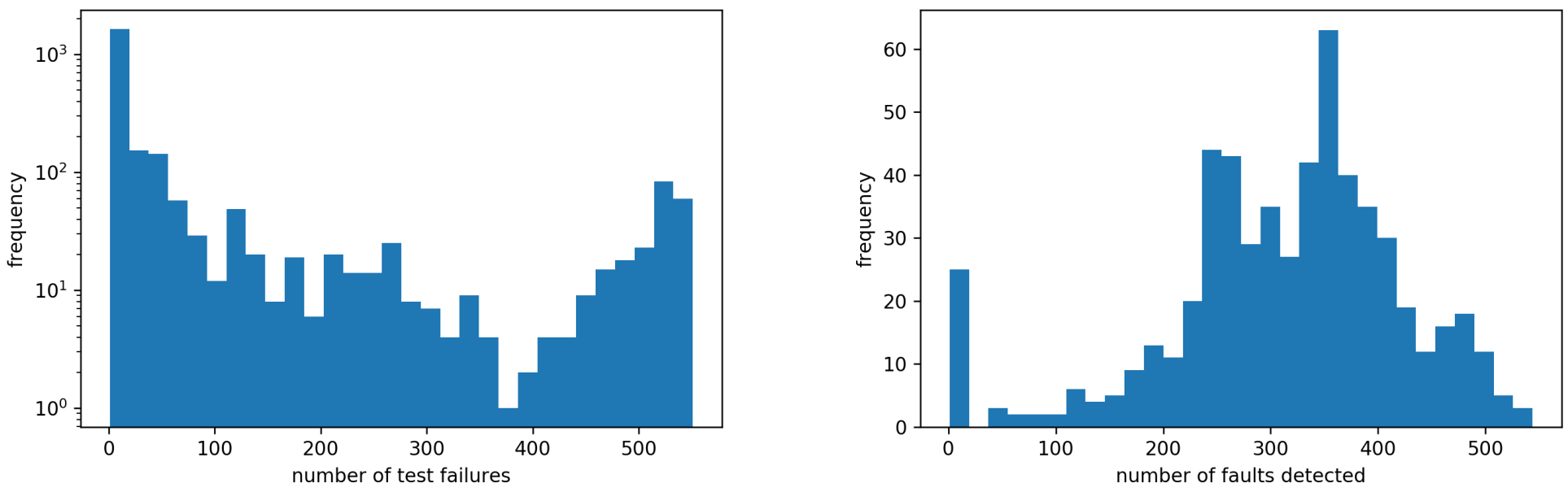

**(d)** Jinja

**Figure 7** Distributions of the number of failing tests of the respective project in each run (left) and the number of seeded defects detected by each test (right).

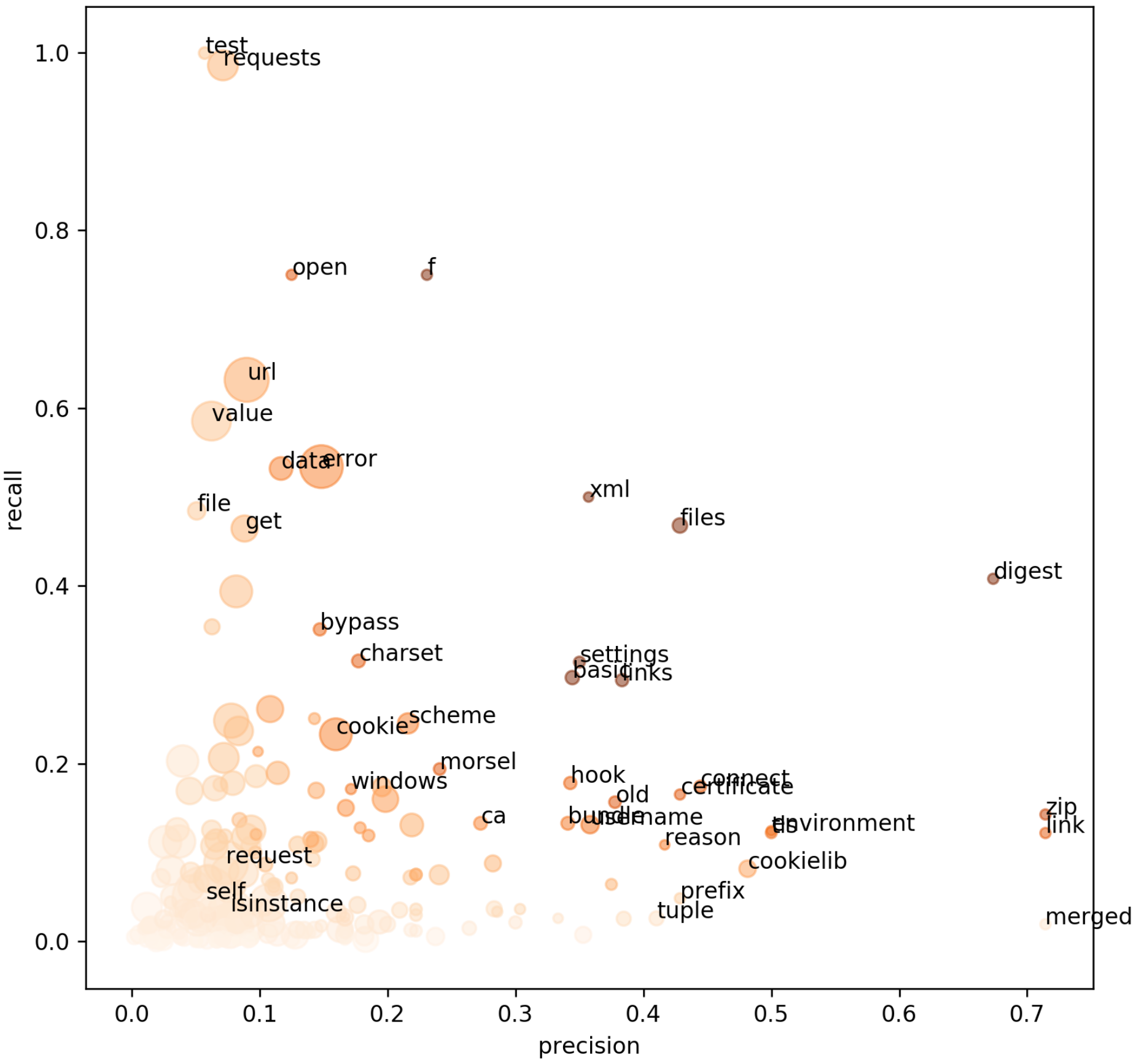

**Figure 8** Distribution of lexical features in the Requests project by precision and recall with respect to predicting failing tests. The darker the colour, the higher their $F_1$ score. The size of a point is proportional to its frequency in the code base.

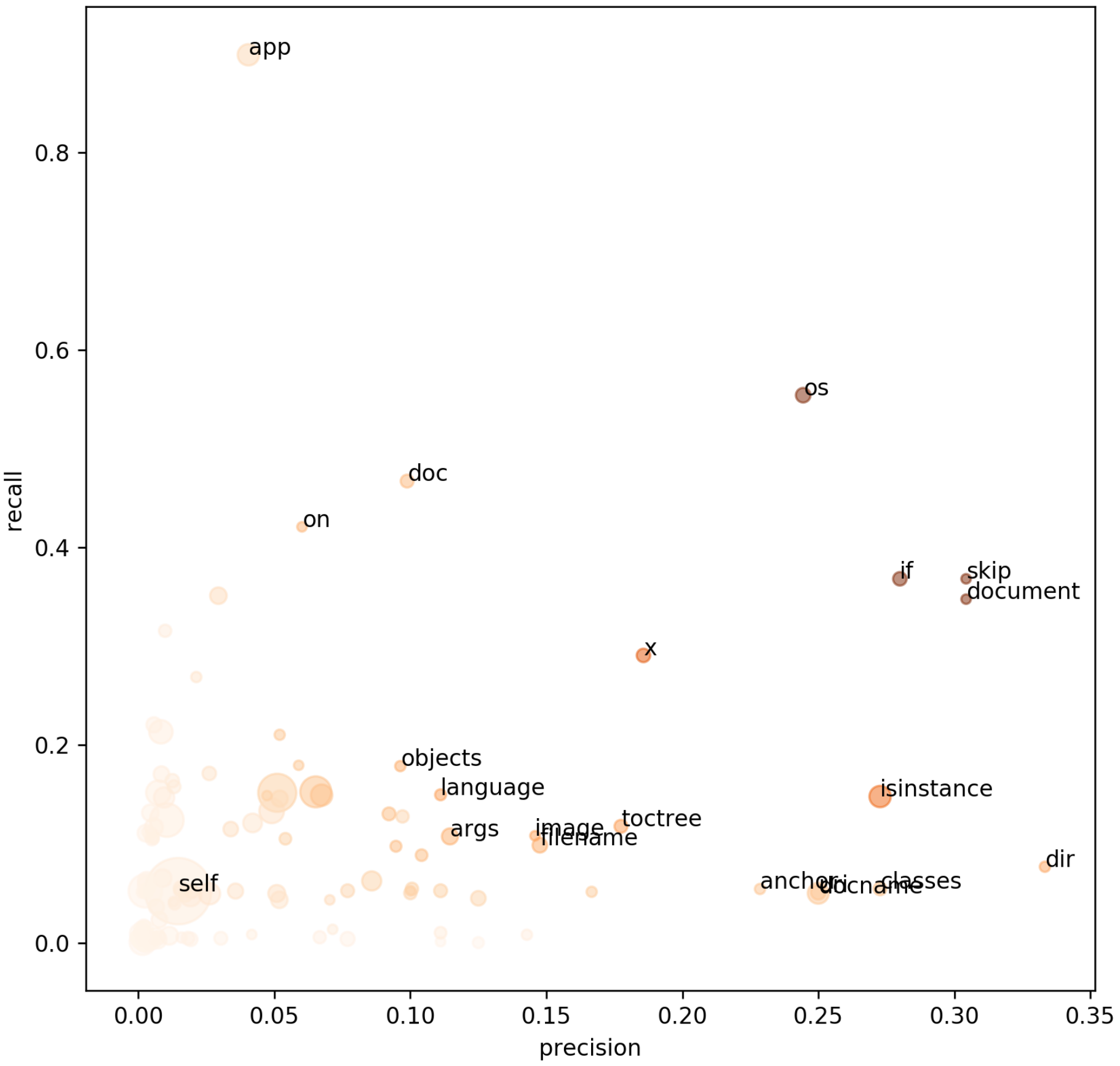

**Figure 9** Distribution of lexical features in the Sphinx project by precision and recall with respect to predicting failing tests. The darker the colour, the higher their $F_1$ score. The size of a point is proportional to its frequency in the code base.

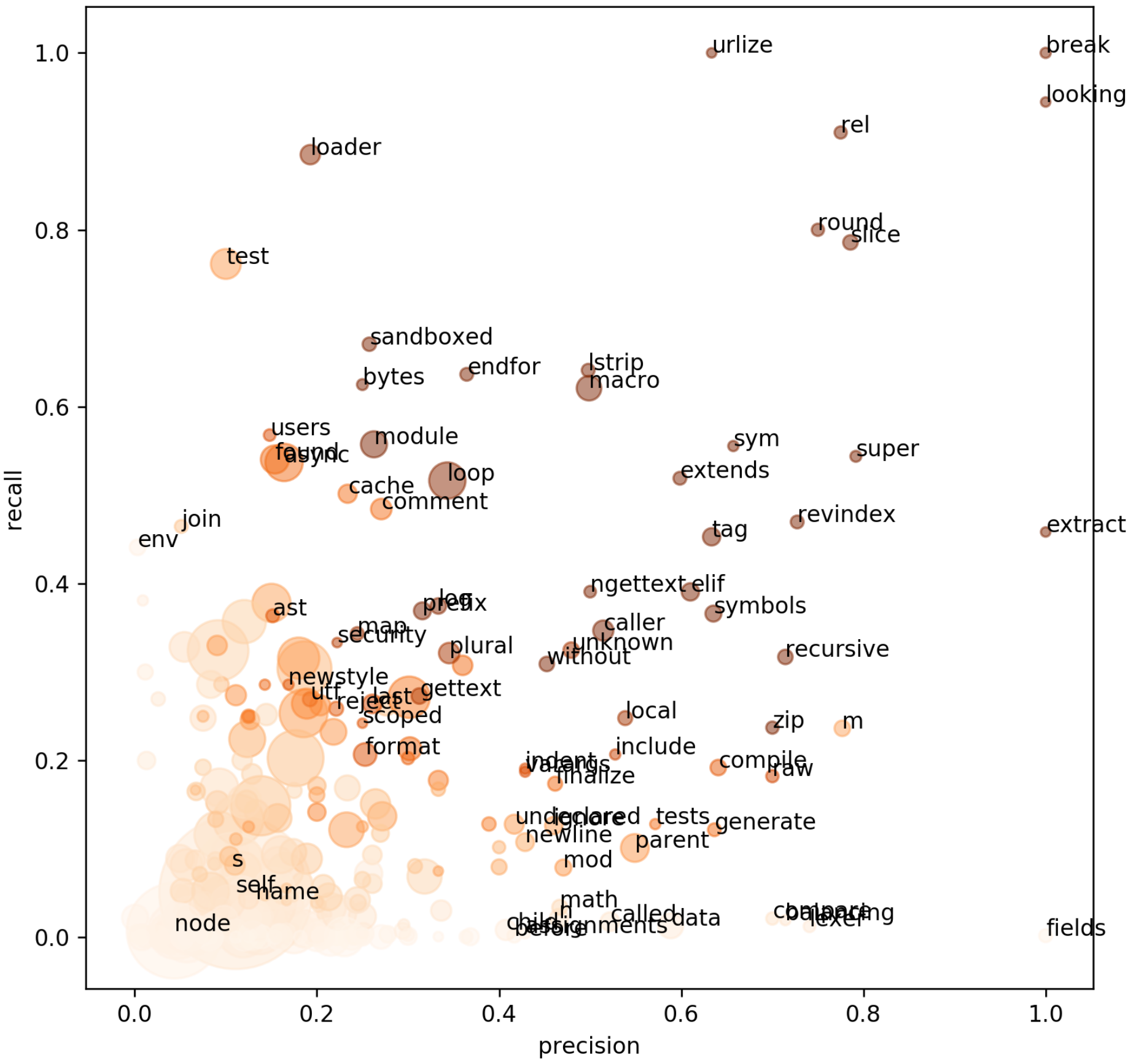

■ **Figure 10** Distribution of lexical features in the Jinja project by precision and recall with respect to predicting failing tests. The darker the colour, the higher their $F_1$ score. The size of a point is proportional to its frequency in the code base.

## About the authors

**Toni Mattis** is a doctoral researcher at the Software Architecture Group. His research interests are software modularity, machine learning for live programming environments, and code repository mining. (toni.mattis@hpi.uni-potsdam.de)

**Robert Hirschfeld** leads the Software Architecture Group at the Hasso Plattner Institute at the University of Potsdam. His research interests include dynamic programming languages, development tools, and runtime environments to make live, exploratory programming more approachable. Hirschfeld received a PhD in computer science from Technische Universität Ilmenau. (robert.hirschfeld@hpi.uni-potsdam.de)